# Approximate discrete dynamics of EMG signal

**Sayan Mukherjee[1], Sanjay Kumar Palit[2], Dilip Kumar Bhattacharya[3]**

[1] Mathematics Department, Shivanath Shastri College, 23/49 Gariahat Road, Kolkata-700029, INDIA.

[2] Mathematics Department, Calcutta Institute of Engineering and Management, 24/1A Chandi Ghosh Road, Kolkata-700040, INDIA.

[3] Rabindra Bharati University, Kolkata-700050, INDIA.

**Abstract.**

Approximation of a continuous dynamics by discrete dynamics in the form of Poincare map is one of the fascinating mathematical tool which can describe the approximate behaviour of the dynamics of the dynamical system in lesser dimension than the embedding diemnsion. The present article considers a very rare biomedical signal like Electromyography (EMG) signal. It determines suitable time delay and reconstruct the attractor of embedding diemnsion three. By measuring its Lyapunov exponent, the attractor so reconstructed is found to be chaotic. Naturally the Poincaré map obtained by corresponding Poincaré section is to be chaotic too. This may be verified by calculation of Lyapunov exponent of the map. The main objective of this article is to show that Poincaré map exists in this case as a 2D map for a suitable Poincaré section only. In fact, the article considers two Poincaré sections of the attractor for construction of the Poincaré map. It is seen that one such map is chaotic but the other one is not so – both are verified by calculation of Lyapunov exponent of the map.

**1. Introduction.**

Learning dynamics from a deterministic nonlinear signal is one of the challenging topics in nonlinear world. Deterministic signal means a signal which is not stochastic and nonlinear signal means a signal whose linear statistical measures convey meaningless informations. Determinsim and nonlinearity of a signal is generally tested by well known Surrogate data method, proposed by Theiler et al. (1992)[1,2]. Now to explain the dyanmics of a signal, phase space reconstruction [3-9] is done by which we can get many information about the dyanmics. Phase space [3-9] is an abstract mathematical concept in Euclidean space which consists of independent coordinates, where the time is absent. Reconstruction means construction of phase space [3-9] from a single observation/signal with the help of suitable time lag and the proper embedding dimension [9-12]. Suitable time lag is such a value from which we can understand after what time the time series components is independent. On the other hand embedding dimension relay -how many independent coordinates are necessary for the reconstruction of phase space [3-9]. Suitable time lag for nonlinear signal is generally found out by Average Mutual Information method (AMI) [13,14] and proper embedding diemnsion is calculated by False Nearest Neighbourhood (FNN) method [9-11]. Any way whenever the process of reconstruction is completed, sometimes it is seen that phase space [3-9] shows irratic behaviour in bounded region due to complex interaction of the independent varaibles. In this situation, dynamics become unpredictable or chaotic. The main reason is that system gradually forget the memory from its initial state. Geometrically it is observed that when distance between two trajectories corresponding to two initial states increases exponentially, phase space [3-9] loses its memory that results in chaos. The measure known as Largest Lyapunov exponent (LLE) [15-17 ] is very useful in this context. In the chaotic phase space [3-9], the most beautyful fact is that how much we observe the chaotic dyanmics through phase space [3-9], we see that trajectories have a tendency to converge on a dense bounded regions (irregular geometrical shaped) but they never intersect with own. Such type of fact is known as deterministic chaos and the dense bounded region is called chaotic attractor. When the attractor is reconstructed from the real world signal, we do not have any knowledge of the long term dynamics of the signal as such. In fact, we do not know the realtion connecting the present position with the previous or past posituions on the moving trajectory. What we can do is that we can try to search for the dynamics on the attractor itself. The dynamics of a high dimensional flow in the corresponding phase space [3-9] is understood conventionally by observing the dynamics induced by the flow on a particular section of the phase space [3-9]. The chosen section, called the Poincaré section [9] helps in visualizing the underlying dynamics. The successive intersections of the flow with the section produce a discrete map known as the Poincaré map [18-25].

The electrical activity of skeletal muscles is reflected in EMG signal [26], which contains information about the structure and function of muscles that make the movement of different parts of the body. The EMG signal [26] conveys information about the controller function of the central and peripheral nervous systems on the muscles. As such, the EMG signal [26] provides a highly useful characterization of the neuromuscular system since many pathological processes, whether arising in the nervous system or the muscle, are manifested by alterations in the signal properties. However, the proper dynamics behind the generation of the EMG signal is still unknown. In fact, if one tries to model the system that generates the EMG signal, the outcome of the model far deviates from the actual outcome (EMG signal). So we cannot rely on the model and hence further mathematical study on EMG signal remains impossible.

In this article, an attempt has been made to understand the dynamics behind the generation of the EMG signal by reconstructing its 3D attractor. As in the present case, the 3D attractor possesses the chaotic structure; we do not calculate the proper embedding dimension for the EMG signal. This is because of the fact that if the 3D attractor is chaotic then it remains so, when it is reconstructed in its actual embedding space. Thus, the Poincaré map obtained in this case will be of two dimension [27] and so the interpretation regarding the dynamics becomes much easier. In fact, our main goal is to find the Poincaré map [18-25] for a suitable Poincare section of the reconstructed attractor, which gives the discrete dyanmics approximation of the continuous EMG signal [26].

## 2. Methodology

### 2.1. Data collection

EMG signal [26] (with noise) in analogue form of the experimental subjects with tremor were recorded in lead-1 and lead-2 configuratuions and collected in data accusation device available in the School of BioScience and Engineering, Jadavpur University, Kolkata 32, India, in which it is converted to digital form. This digitized data was then processed in a laptop by using LAB VIEW software to remove noise. Finally, the recorded data was analyzed using a MATLAB program.

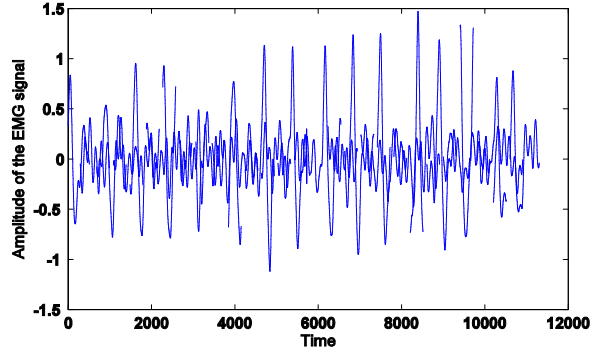

**Figure.1. EMG signal with 11000 samples.**

## 2.2. Phase space reconstruction

Taken theorem states that it is possible to reconstruct a topological equivalent phase space [3-9] from a single time series [28, 29], if the suitable time delay and proper embedding dimension can be found out. Let us consider the time series data given by $\{x(k), k = 1, 2, \ldots, N\}$. Suppose the embedding dimension and the delay time for reconstruction of the attractor are $m$ and $\tau$ respectively. Then reconstructed phase space [3-9] is given by $X(k) = (x(k), x(k+\tau), x(k+2\tau), \ldots, x(k+(m-1)\tau))$, $k = 1, 2, \ldots, M$, where $\{x(k)\}$ is the phase space's point in $m$-dimension phase space [3-9], $M$ is the number of phase points, $M = N - (m-1)\tau$, describes the evaluative trajectory of the system in the phase space [3-9]. Reconstruction of the attractor is guaranteed if the dimension of the phase space [3-9] is sufficient to unfold the attractor. It is ensures when $m > 2d + 1$ will achieve this, where $d$ is the dimension of the attractor.

For the nonlinear signal, suitable time delay is calculated by the method of Average mutual information. The mutual information is a measure of how much information can be derived from one point of a time series, given complete information about the other. On the other hand, embedding dimension is calculated by FNN [9-11], but for the sake of geometrical visibility we reconstruct the attractor with embedding dimension three.

### 2.2.1. Average Mutual information method

For a time series $\{x(k), k = 1, 2, \ldots, N\}$, AMI [13, 14] is calculated by

$$AMI(\tau) = \sum_{k=1}^{N-\tau} prob[x(k), x(k+m)] \log \frac{prob[x(k), x(k+m)]}{prob[x(k)]prob[x(k+m)]}, \quad (1)$$

$[m = 1, 2, \ldots, N-1]$ where $prob[\bullet]$ denotes the probability.

For estimation of $\tau$, two criteria are important. First, $\tau$ has to be large enough so that the AMI [13, 14] at time $k + \tau$ is significantly different from the AMI [13,14] at time $k$. Then it will be possible to gather enough information about all other system variables that influence the value of $x$ to reconstruct the whole attractor. Second, $\tau$ should not be larger than the typical time for which the system loses memory of its initial state. We will always conscious about the second criteria, because chaotic systems are unpredictable or lose memory of its initial state as time goes forward. In this context, Fraser and

Swinney proposed a very useful method [13, 14] which state that the optimum time-delay is obtained where the mutual information attains its first minimum value.

### 2.3. Nonstationarity test

Nonstationary signal means a signal where the pattern of probabilty distributions of different segments are not equal at all. We simple use Quantile-Quantile plot (QQ-plot) [30] to test the nonstationarity of EMG signal [26]. Basically it is a graphical technique for determining if two data sets come from populations with a common distribution. For this purpose, a 45-degree reference line is plotted. If the two sets come from a population with the same distribution, the points should fall approximately along this reference line. The greater the departure from this reference line, the greater the evidence for the conclusion that the two data sets have come from populations with different distributions.

### 2.4. Nonlinearity by Surrogate data Test (with 99% confidence)

Surrogate data of an observed signal is such a time series which has same linear statistical properties as in the observed signal. Surrogate data Test method is based on statistical hypothesis testing [30]. This method has three steps: generate 99 surrogate data from the observed data. In this paper, we observed that EMG signal [26] is a non-stationary signal, so Surrogate data is thus generated by Amplitude Adjusted Truncated Fourier Transform surrogates (AATFT) method [31] and select the nonlinear version of autocorrelation statistics-AMI (with m=1) as discriminate statistics. Consider a null hypothesis against which observations are tested. Here we consider the null hypothesis ($H_0$) as
$H_0 : AMI_{\text{experimental signal}}(m=1) = AMI_{SUR(\text{experimental signal})}(m=1)$.

Here AMI [13, 14] plays a role as the discriminating statistic, which is basically a number that quantifies some aspect of the time series. If this number is different for the observed data then null hypothesis can be rejected.

### 2.5. Largest Lyapunov exponent and signature of chaos

Detecting the presence of chaos in a dynamical system is an important problem that is solved by measuring the LLE [15-17]. Lyapunov exponents [15-17] quantify the exponential divergence of initially close state-space trajectories and estimate the amount of chaos in a system. We use a new method for calculating the LLE [15-17] from an experimental time series. The method follows directly from the definition of the LLE [15-17] and is accurate because it takes advantage of all the available data. The major advantages of this algorithm are that the algorithm is fast, easy to implement, and robust to changes in the following quantities: embedding dimension, size of data set, reconstruction delay, and noise level.

The LLE [15-17] is easily and accurately calculated using a least-square fit to the "average" line defined by

$$y(i) = \frac{1}{\Delta t} \langle \ln d_j(i) \rangle, \text{ where} \qquad (2)$$

$\langle . \rangle$ denotes the average over all values of $j$ and $d_j(i)$ is the distance between the $j$-th pair of nearest neighbours after $i$ discrete time-steps, i.e., $i\Delta t$ seconds.

## 2.6. Poincaré section and Poincaré map

Let $P = \{(x(k), x(k+m), x(k+2m)) \in R^3\}_{m=1}^{N-m}$ be a 3D phase space [3-9] and z= c be a plane which cut the phase space [3-9] orthogonally. For the experimental data set, the first task is to find out those points from phase space [3-9] which is very near to the plane z=c in one side. Let they are $(x(h), x(h+m), x(h+2m))$ for $h = a_1, a_2, a_3, ..., a_N$, where $a_i (i=1,2,..,N)$ denotes the time index and $a_{i+1} - a_i \neq a_{i+2} - a_{i+1} (i=1,2,..,N-2)$. Then, find out $\{(x(h+1), x(h+1+m), x(h+1+2m))\}$ which are just near on the other side of the plane. Joining $\{(x(h), x(h+m), x(h+2m))\}$ to $\{(x(h+1), x(h+1+m), x(h+1+2m))\}$ by a line which cuts the plane z=c orthogonally. Suppose the joining lines meets the plane z=c at $(u(n+1), v(n+1)), n = 1,2,...$. The plane z=c, containing those points, is known as Poincaré section [9]. Next we draw two surfaces $u(n+1) = f(u(n), v(n)), v(n+1) = g(u(n), v(n)), n = 1, 2, 3,..$ and find the intersections of them. This intersection gives us a three dimensional curve called Poincaré map [18-25], since it is described by 2D map: $u(n+1) = f(u(n), v(n)), v(n+1) = g(u(n), v(n)), n = 1, 2, 3,..$.

## 2.7. Lyapunov exponent of a map

Consider a map $u(n+1) = f(u(n), v(n)), v(n+1) = g(u(n), v(n)), n = 1, 2, 3,.., N$. Then the partial differences $\Delta_u f, \Delta_v f, \Delta_u g, \Delta_v g$ are given by

$$\Delta_u f = f(u(n+1), v(n)) - f(u(n), v(n)),$$
$$\Delta_v f = f(u(n), v(n+1)) - f(u(n), v(n)),$$
$$\Delta_u g = g(u(n+1), v(n)) - g(u(n), v(n)),$$
$$\Delta_v g = g(u(n), v(n+1)) - g(u(n), v(n)).$$

Let $J = \begin{pmatrix} \Delta_u f & \Delta_v f \\ \Delta_u g & \Delta_v g \end{pmatrix}$.

Thus, for each of $N$ - pairs of $(u(n), v(n))$, we get $N$ $2 \times 2$ matrices which are given by $J_1, J_2, ......, J_N$. Then $G = J_1 \cdot J_2 \cdot ...... \cdot J_N$ is also a $2 \times 2$ matrix. Suppose $r_k, k = 1, 2.$ are the Eigen values of $G$. Then, Lyapunov Exponent is defined by

$$\}_k = \frac{1}{N} \log(r_k), k = 1, 2.$$

If $r_k (k=1,2)$ are complex conjugates, then $\}_k (k=1,2)$ are complex. If any one of $r_k (k=1,2)$ is negative, then the corresponding value of $\}_k$ is also complex. In both the cases $\}_k$ is not workable, as complex Lyapunov exponents do not signify exponential divergence. So, the only case to be considered is when both $r_k$ are real and positive. Further if one of $\}_k (k = 1, 2)$ is positive and the other one is negative, then it ensures that the map is chaotic.

## 3. Result and discussion

### 3.1. Nonstationarity of the EMG signal

The QQ-plot of the given EMG signal is shown by figure. 2.

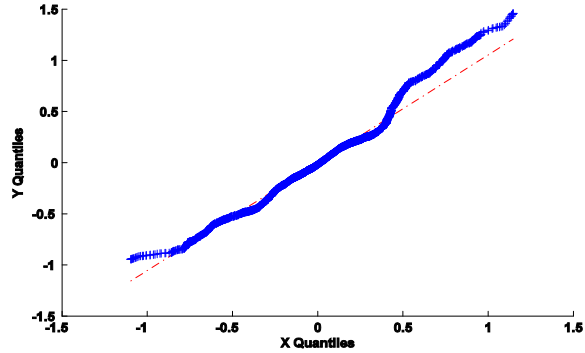

**Fig.2. QQ-plot of EMG signal shows many points lie far away from the diagonal line *y=x*.**

From the above QQ-plot [29] it is evident that the distributions are different for any two segments of equal length. This proves that the statistical parameters of the segments of equal length are always different. In other words, the EMG signal [26] is non-stationary.

### 3.2. Nonlinearity of EMG signal

After generating 99 surrogate data of the EMG, we calculate AMI (with m=1) for each of them and then draw a graph Grade vs. AMI (with m=1), which is given in figure 2

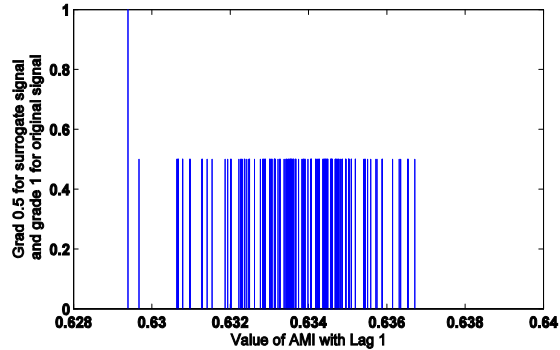

**Figure. 3. Two grades- '1' and '0' are considered in y-axis. For the collected data we considered '1' and for the surrogate data we considered '0'. In y-axis, we fix the grade '1' for the EMG signal and 0.5 for the surrogate data. In x-axis, we take the value of the AMI (with lag=1).**

From figure 2, it is seen that value of AMI (with m=1) of EMG signal [26] is different of the values of the AMI (with lag=1) of each 99 surrogate data.
We consider the null hypothesis ($H_0$) as follows:

$H_0$: $AMI_{EMG}(m=1) = AMI_{SUR(EMG)}(m=1)$ where $AMI_{EMG}$ denotes *AMI* of EMG and $AMI_{SUR(EMG)}$ denotes

AMI [13, 14] of surrogate data of EMG. Hence by the method of surrogate data test [30], null hypothesis is rejected with 99% confidence. So, that EMG is a nonlinear time series.

### 3.3. 3D chaotic phase space reconstruction of EMG signal

### 3.3.1. Time-delay

AMI vs. time-delay graph is given in figure 3. Since first minimum value of AMI [13, 14] occurs at lag=33, so the value of time delay / lag is 33. This method actually determines the time lag by which we can get independent coordinates from a single time series for the attractor reconstruction.

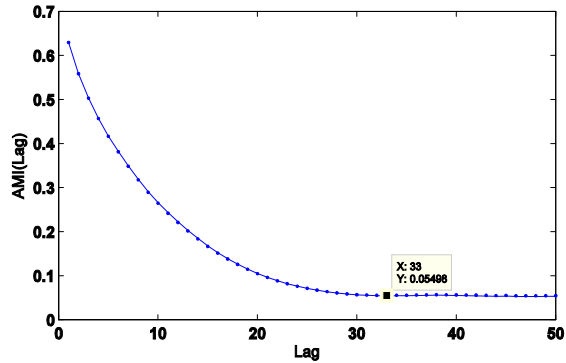

**Figure. 4. X-axis represents time delay/lag and y-axis represents AMI. AMI has the first minimum value (=0.5498) when the time-lag (m)=33.**

### 3.3.2. 3D Phase-space

3D reconstructed phase space [3-9] with m=33 of the EMG signal [26] is given by, $X(k) = (x(k), x(k+33), x(k+66))$, where $k = 1, 2, ....., N - 66$. The following figure shows the 3D reconstructed phase space [3-9]:

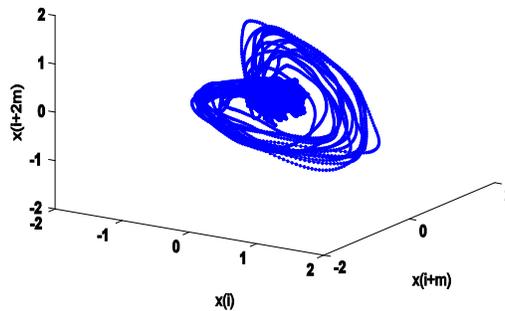

**Figure. 5. 3D reconstructed attractor with time lag ($m$) =33.**

## 3.4. Poincaré section and 2D Poincaré map

Figure.6a and figure.6b shows the Poincaré section at $z = 0.174$ and at $z = 0.1123$ respectively. The dots in the XY-plane are those points, where the trajectories meet $z = 0.174$ and $z = 0.1123$.

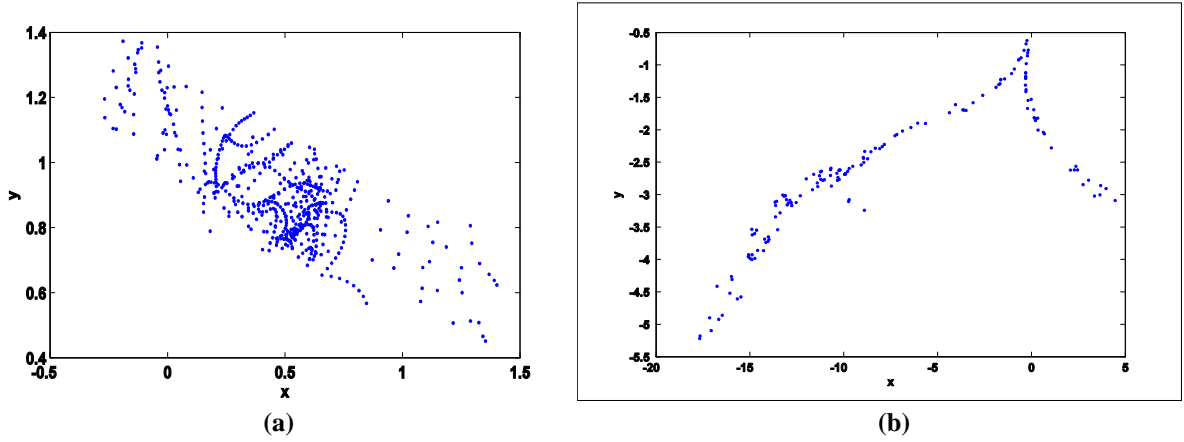

(a)        (b)

**Figure 6. 2D Poincaré section of the 3D reconstructed attractor (a) at $z = 0.174$ and (b) at $z = 0.1123$.**

It is observed from the above diagram that the points on the Poincaré section are mixed in the sense that this diagram does not carry the information about the times at which those points occur. So, from this collection of dots (points), we first find out the succesive sequence of dots in order of their arrival on the Poincaré section. Let us call the rearranged sequence of dots as $(u(n), v(n))$ and make two triplets – $(u(n), v(n), u(n+1))$ and $(u(n), v(n), v(n+1))$. Next plot these triplets separately in 3D and fit surfaces individually. For the section $z = 0.174$, surfaces are shown in figure. 7a and figure.7b.

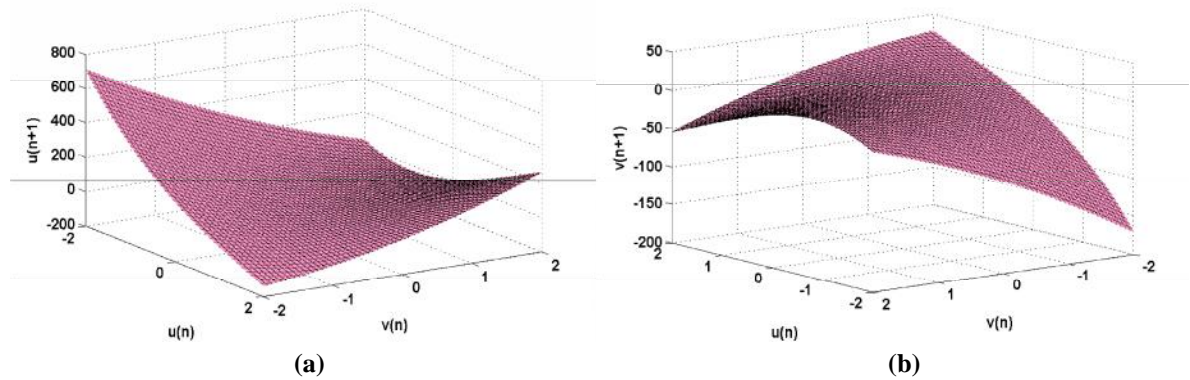

(a)        (b)

**Figure 7. (a) Fitted surface $u(n+1) = f(u(n), v(n))$, (b) Fitted surface $v(n+1) = f(u(n), v(n))$**

It is observed that the functional form of the fitted surface $u(n+1) = f(u(n), v(n))$ is given by

$$u(n+1) = 20.27 - 48.85u(n) - 34.03v(n) + 35.02u^2(n) + 67.13u(n)v(n) + 14.22v^2(n) - 7.501u^3(n) - 25.4u^2(n)v(n) - 21.2u(n)v^2(n). \quad (3)$$

Also we observed that the functional form of the fitted surface $v(n+1) = f(u(n), v(n))$ is given by

$$v(n+1) = -4.971 + 11.78u(n) + 9.145v(n) - 8.777u^2(n) - 14.83u(n)v(n) - 3.363v^2(n) +$$
$$2.093u^3(n) + 6.044u^2(n)v(n) + 4.277u(n)v^2(n) \tag{4}$$

For the section $z = 0.1123$, surfaces are shown in figure. 8a and figure.8b.

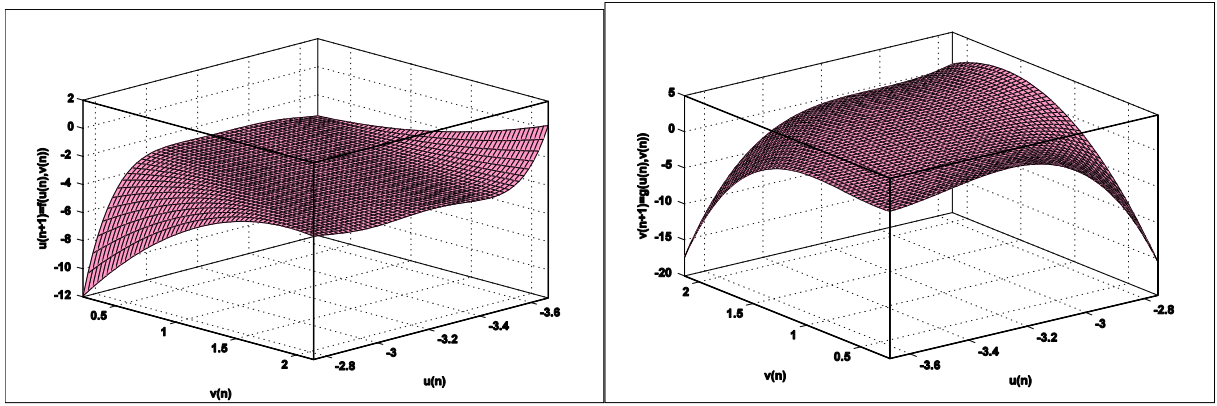

**(a)**                  **(b)**

**Figure 8.** (a) Fitted surface $u(n+1) = f(u(n), v(n))$, (b) Fitted surface $v(n+1) = f(u(n), v(n))$.

It is observed that the functional form of the fitted surface $u(n+1) = f(u(n), v(n))$ is given by

$$u(n+1) = -6.692e+004 - 9.893e+004u(n) + 1.365e+004v(n)$$
$$- 5.843e+004u(n)^2 + 1.637e+004u(n)v(n) - 657.7v(n)^2$$
$$- 1.723e+004u(n)^3 + 7358u(n)^2v(n) - 595.4u(n)v(n)^2 \tag{5}$$
$$- 2536u(n)^4 + 1469u(n)^3v(n) - 179.6u(n)^2v(n)^2 - 149u(n)^5$$
$$+ 109.9u(n)^4v(n) - 18.05u(n)^3v(n)^2.$$

Also we observed that the functional form of the fitted surface $v(n+1) = f(u(n), v(n))$ is given by

$$v(n+1) = -5207 - 5226u(n) + 3440v(n) - 1924u(n)^2 + 2830u(n)v(n)$$
$$- 508.2v(n)^2 - 306.3u(n)^3 + 766.4u(n)^2v(n) - 299.6u(n)v(n)^2$$
$$+ 19v(n)^3 - 17.66u(n)^4 + 68.44u(n)^3v(n) - 43.03u(n)^2v(n)^2 \tag{6}$$
$$+ 6.832u(n)v(n)^3.$$

The intersection of the surfaces described in (3) & (4) and the intersection of the surfaces described in (5) & (4) are shown in figure.9a and figure.9b respectively.

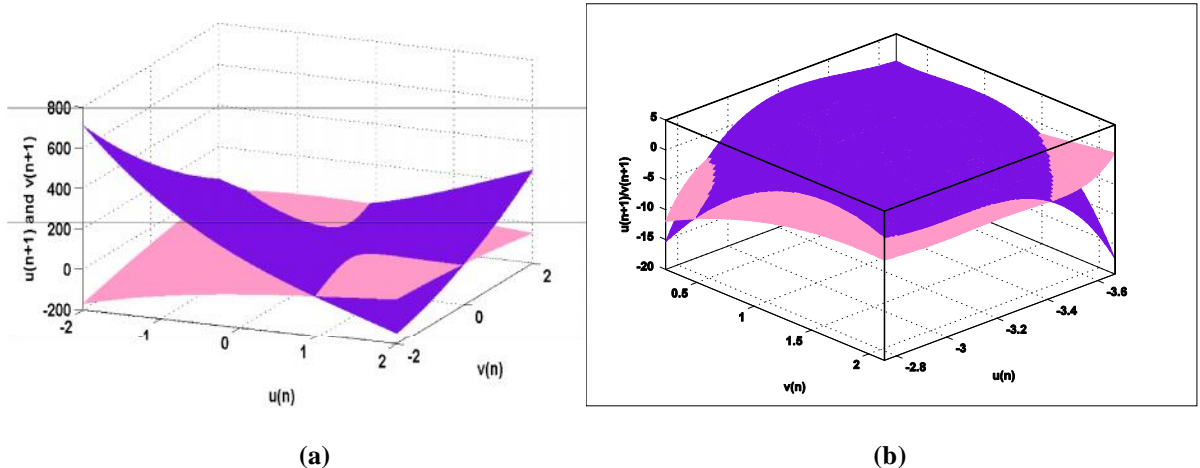

**(a)** **(b)**
**Figure 9. (a) Intersection of the two surfaces (3) &(4), (b) Intersection of the two surfaces (5) &(6) .**

Intersection shows a 3D curve which is known as 2D Poincaré map of the reconstructed attractor of the EMG signal [26]. Thus, it is established that 2D Poincaré map in form of difference equation exists for a nonlinear EMG signal [26].

### 3.5. Verification of proper Poincaré map

Since there may exists many Poincaré sections for the reconstructed attractor of EMG signal, so there may be many Poincaré maps, one for each Poincaré section. Now the natural question arises – which Poincaré sections are perfect among them and which Poincaré map we choose? Otherwise we cannot say that approximation of continuous dynamics is complete at all. To answer this query, we calculate Lyapunov exponent [32] for each of the Poincaré maps of the corresponding Poincaré section.

Let $u(n+1) = f(u(n), v(n))$ and $v(n+1) = f(u(n), v(n))$ be two surfaces whose intersection is the required 2D Poincaré map. Next let $(p(n), q(n), r(n))$ denotes the points in 3D such that $\{(p(n), q(n), r(n))\} = \{(u(n), v(n), u(n+1)) : u(n+1) = f(u(n), v(n))\} \cap \{(u(n), v(n), v(n+1)) : v(n+1) = g(u(n), v(n))\}$.

Figure.10a and figure.10b show such points in 3D for the Poincaré section $z = 0.174$ and $z = 0.1123$ respectively.

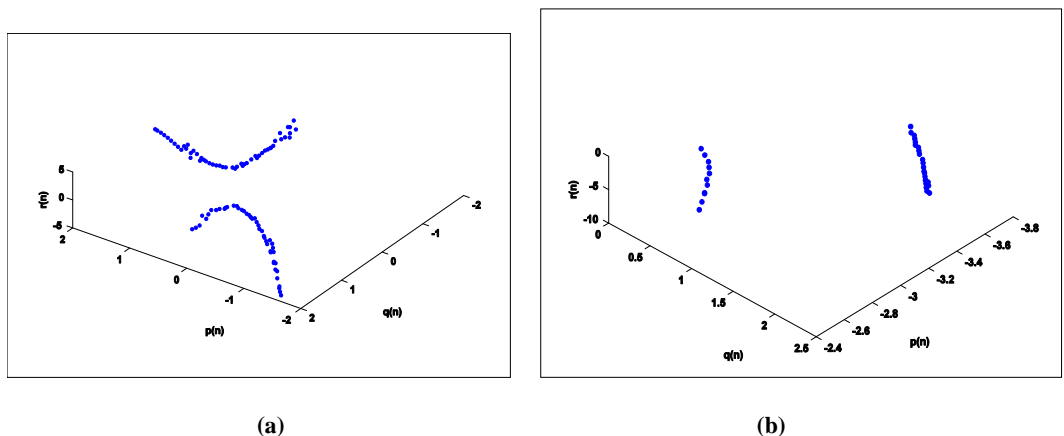

**(a)** **(b)**
**Figure.10. (a) Points of intersection of the two surfaces given by (3) and (4), (b) Points of intersection of the two surfaces given by (5) and (6).**

Now, to calculate the Lyapunov exponent of the 2D Poincaré map, we consider only two variables – $u(n)$ and $v(n)$, because the maps (given by (3), (4) and also by (5), (6)) are described by only those two variables. Thus the third variable has no role to play in this case and hence we take the 2D projection of the above Poincaré maps for finding the Lyapunov exponent. These are shown by figure.11a and figure.11b respectively.

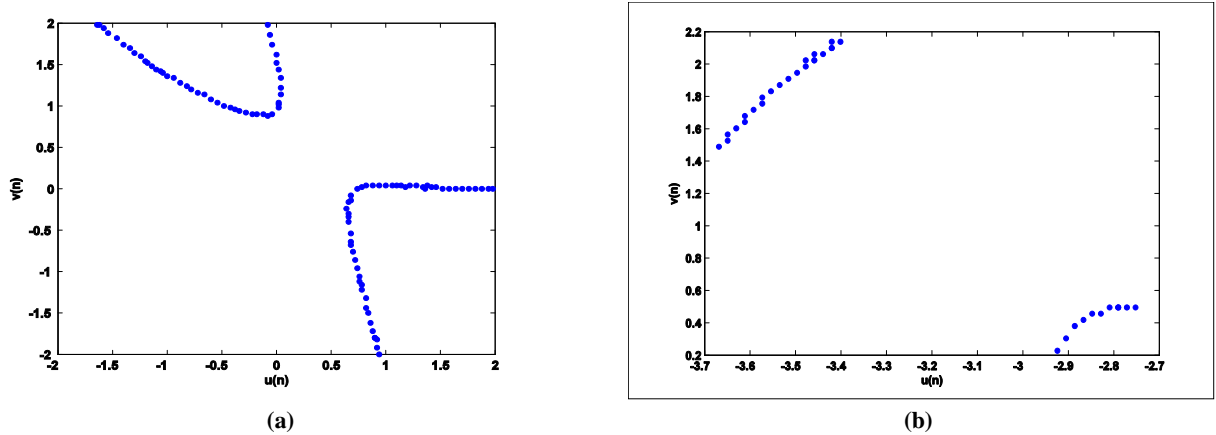

(a)            (b)

**Figure.11. 2D Projections of (a) the Poincaré map given by figure.10a, (b) the Poincaré map given by figure.10b.**

The resultant matrices for the Poincaré section at $z = 0.174$ and $z = 0.1123$ are given by

$$G_{z=0.174} = \begin{pmatrix} 2.2605 & -0.1205 \\ -2.6244 & 0.1689 \end{pmatrix} \text{ and } G_{z=0.1123} = \begin{pmatrix} -0.1159 & 0.0506 \\ -0.0933 & 0.0407 \end{pmatrix}$$

Let the two Eigen values of $G_{z=0.174}$ and $G_{z=0.1123}$ be $r_{1,z=0.174}, r_{2,z=0.174}$ and $r_{1,z=0.1123}, r_{2,z=0.1123}$ respectively. The calculated value of $r_{1,z=0.174}, r_{2,z=0.174}; r_{1,z=0.1123}, r_{2,z=0.1123}$ are given by

$r_{1,z=0.174} = 0.0127, r_{2,z=0.174} = -0.0522, r_{1,z=0.1123} = -0.0752$ and $r_{2,z=0.1123} = 6.9389\text{e-}018$.

Next let the corresponding Lyapunov exponents be $\}_{1,z=0.174}, \}_{2,z=0.174}$ and $\}_{1,z=0.1123}, \}_{2,z=0.1123}$ respectively. The calculated value of $\}_{1,z=0.174}, \}_{2,z=0.174}$, $\}_{1,z=0.1123}, \}_{2,z=0.1123}$ are given by

$\}_{1,z=0.174} = 0.0127, \}_{2,z=0.174} = -0.0522; \}_{1,z=0.1123} = -0.0761 + 0.0924i, \}_{2,z=0.1123} = -1.1620.$

## 4. Conclusion

Study of continuous chaotic dynamics of some known dynamical model through its discrete dynamics approximation in form of Poincaré map is a very common practice in nonlinear analysis. The purpose of such approximation is to understand the higher dimensional dynamics in lower dimension. It would be better if the similar approximation can be done for any continuous nonlinear signal. This might be helpful to understand the approximate dynamics behind the generation of that signal. However, no such study is available for a continuous nonlinear signal, whose dynamical model is not known. In the present article an attempt has been made to show that the above study can also be done for a nonlinear EMG signal [26] without having any knowledge of its dynamics. In fact, we have reconstructed a proper 3D chaotic attractor from a nonlinear continuous EMG signal under suitable choice of time-delay and successfully approximated the chaotic dynamics of the signal [26] by a chaotic 2D Poincaré map for a suitable Poincaré section of the reconstructed attractor. However, the chaotic nature of the map ceases to exist if the Poincaré section is not suitably chosen. Thus it may be be concluded that the chaotic nature of a signal may be revealed through construction of its chaotic 2D Poincaré map only. In other words, the continuous chaotic dynamics may be approximated by a discrete dyanmics in form of chaotic Poincaré map directly from a continuous real life signal without having any dynamical model behind its generation.

**Refernces**

[1] J. Theiler, S. Eubank, A. Longtin, B. Galdrikian, J. Farmer, Testing for nonlinearity in time series: the method of surrogate data, Physica D 58 (1992) 77–94.

[2] D. Kugiumtzis, Test your surrogate data before you test for nonlinearity, Phys. Rev. E 6 (1999) 2808–2816.

[3] F. Takens, Detecting strange attractors in turbulence In Dynamical systems and turbulence, Lecture Notes in Mathematics 898 (1981) 366–381.

[4] D. T. Kaplan, L. Glass, Understanding Nonlinear Dynamics, Springer, New York, 1995.

[5] S. H. Strogatz, Nonlinear Dynamics and Chaos, Addison-Wesley, 1994.

[6] E. Ott, Chaos in dynamical systems, Cambridge University Press, 1993.

[7] D. Ruelle, Chaotic Evolution and Strange Attractors, Cambridge University Press, 1989.

[8] J. L. McCauley, Chaos, Dynamics, and Fractals: An Algorithmic Approach to Deterministic Chaos, Cambridge University Press, 1993.

[9] G. P. Williams, Chaos Theory Tamed, Joseph Henry Press, Washington D.C., 1997.

[10] M. B. Kennel, R. Brown , H. D. I. Abarbanel, Determining embedding dimension for phase-space reconstruction using a geometrical construction. Phys. Rev. A, 45(6) (1992) 3403-3411.

[11] W. Liebert, K. Pawelzik, H. G. Schuster, Optimal embeddings of chaotic attractors from topological considerations. Eur. phys. Lett. 14(6) (1991) 521-526.

[12] H. Pi, C. Peterson, Finding the embedding dimension and variable dependencies in time series. Neural Comp. 6 (1994) 509-520.

[13] A.M. Fraser, H.L. Swinney, Independent coordinates for strange attractors from mutual information, Phys. Rev. A 33 (1986) 1134–1140.

[14] W. Liebert, H. G. Schuster, Proper choice of time delay for the analysis of chaotic time series. Phys. Lett. A 142 (1989) 107.

[15] K. Briggs, An improved method for estimating Liapunov exponents of chaotic time series. Phy. Lett. A 151 (1990) 27-32.

[16] A. Wolf, B. J. Swift, H. L. Swinney, J. A. Vastano, Determining Lyapunov exponents from a time series, Physica D 16 (1985) 285-317.

[17] Michael T. Rosenstein, James J. Collins and Carlo J. De Luca, A practical method for calculating largest Lyapunov exponents from small data sets, Physica D 65 (1993) 117-134.

[18] S. H. M. J. Houben, J. M. L. Maubach, R. M. M. Mattheij, An accelerated Poincaré -map method for autonomous oscillators, Appl. Math. Comp. 140 (2003) 2-3.

[19] W. Just, H. Kantz, Some considerations on Poincaré maps for chaotic flows, J. Phys. A: Math. Gen. 33 (2000)163.

[20] T. Parker, L. Chua, Practical numerical algorithms for chaotic systems, Springer, New York, 2000.

[21] H. Poincaré, Les méthodes nouvelles de la méchanique céleste, Gauthier-Villars, Paris, 1899.

[22] H. G. Schuster, Deterministic chaos, VCH, Weinheim, 1989.

[23] R. Efrem, Numerical approximation of Poincaré maps, Romai. J. 4 (2008)101-106.

[24] P. Cvitanovi, R. Artuso, R. Mainieri, G. Tanner, G. Vattay, Chaos: Classical and Quantum, Niels Bohr Institute, Copenhagen, 2007.

[25 ] M. Henon, On the numerical computation of Poincaré maps, J. Physica D 5(2-3) (1982) 412-414.

[26] K. J. Blinowska, J. Zygierewicz, Practical Biomedical Signal Analysis Using MATLAB®, CRC Press, Taylor & Francis Group, 2012.

[27] S. Mukherjee, S. K. Palit , D.K. Bhattacharya, Is one dimensional Poincaré map sufficient to describe the chaotic dynamics of a three dimensional system?, Applied Mathematics and Computation 219, (2013) 11056–11064.

[28] J.P. Huke , Embedding Non-Linear Dynamical Systems: A Guide to Takens Theorem, Manchester Ins. of Math. Sci. 26, 2006.

[29] F.Takens, Detecting strange attractors in turbulence, Lecture Notes in Mathematics 898, (1981) 366–381.

[30] W. L. Martinez and A. R. Martinez, Computational Statistics Handbook with Matlab, Chapman & Hall/CRC, 2002.

[31] H. Kantz, T. Schreiber, Nonlinear Time Series Analysis, Cambridge University Press, 2004, pp.334.

[32] N. Das, Determination of Lyapunov Exponents in Discrete Chaotic Models, International Journal of Theoretical & Applied Sciences **4**(2), (2012) 89-94